\documentclass[aps,prl,amssymb,reprint,onecolumn,superscriptaddress,longbibliography]{revtex4-1}

\usepackage[ansinew]{inputenc}
\usepackage{graphicx}
\usepackage[breaklinks=false,colorlinks=true,linkcolor=blue,urlcolor=blue,citecolor=blue]{hyperref}
\usepackage{setspace}
\usepackage[ansinew]{inputenc}
\usepackage{caption}
\usepackage[labelfont=bf]{caption}
\usepackage{graphicx}
\usepackage{amsmath}
\usepackage{color}
\usepackage{lineno}
\usepackage{ulem}
\begin{document}

\title{Attosecond multi-photon multi-electron dynamics} 

\author{M. Kretschmar}
\thanks{These authors contributed equally to this work.}
\affiliation{Max-Born-Institut, Max-Born-Strasse 2A, 12489 Berlin, Germany}
\author{A. Hadjipittas}
\thanks{These authors contributed equally to this work.}
\affiliation{Department of Physics and Astronomy, University College
 London, Gower Street, London WC1E 6BT, United Kingdom}
\author{B. Major}
\affiliation{ELI-ALPS, ELI-HU Non-Profit Ltd., Wolfgang Sandner utca 3., Szeged 6728, Hungary}
\affiliation{Department of Optics and Quantum Electronics, University of Szeged, D\'om t\'er 9, Szeged 6720, Hungary}
\author{J. T\"ummler}
\affiliation{Max-Born-Institut, Max-Born-Strasse 2A, 12489 Berlin, Germany}
\author{I. Will}
\affiliation{Max-Born-Institut, Max-Born-Strasse 2A, 12489 Berlin, Germany}
\author{T. Nagy}
\affiliation{Max-Born-Institut, Max-Born-Strasse 2A, 12489 Berlin, Germany}
\author{M. J. J. Vrakking}
\affiliation{Max-Born-Institut, Max-Born-Strasse 2A, 12489 Berlin, Germany}
\author{A. Emmanouilidou}
\email{a.emmanouilidou@ucl.ac.uk}
\affiliation{Department of Physics and Astronomy, University College London, Gower Street, London WC1E 6BT, United Kingdom}
\author{B. Sch\"utte}
\email{Bernd.Schuette@mbi-berlin.de}
\affiliation{Max-Born-Institut, Max-Born-Strasse 2A, 12489 Berlin, Germany}

\date{\today}

\begin{abstract}
Multi-electron dynamics in atoms and molecules very often occur on sub- to few-femtosecond timescales. The available intensities of extreme-ultraviolet (XUV) attosecond pulses have previously only allowed the time-resolved investigation of two-photon, two-electron interactions. Here we demonstrate attosecond control over double and triple ionization of argon atoms involving the absorption of up to five XUV photons. In an XUV-pump XUV-probe measurement using a pair of attosecond pulse trains (APTs), the Ar$^{2+}$ ion yield exhibits a weak delay dependence, suggesting that its generation predominantly results from the sequential emission of two electrons by photoabsorption from the two APTs. In contrast, the Ar$^{3+}$ ion yield exhibits strong modulations as a function of the delay, which is a signature of the simultaneous absorption of at least two XUV photons. The experimental results are well reproduced by numerical calculations that provide detailed insight into the ionization dynamics. Our results open up new opportunities for the investigation and control of multi-electron dynamics and complex electron correlation mechanisms on extremely short timescales.
\end{abstract}
\maketitle

Strong laser fields can be used to induce multi-electron dynamics in atoms and molecules~\cite{mainfray91, brabec00}. For example, in the commonly used near-infrared (NIR) regime, techniques like high-harmonic generation (HHG) spectroscopy have helped to identify multi-electron processes, such as the rearrangement of electrons in molecules upon photoionization~\cite{smirnova09} and collective multi-electron effects in atoms~\cite{shiner11}. However, the use of NIR fields comes at the cost of strongly perturbing the system and involves a strong bending of the atomic and molecular potentials. The high complexity of the contributing processes then often makes a detailed understanding of the underlying physics difficult. A better control over multi-electron dynamics can be achieved by ionizing atoms and molecules using XUV photons, for which ponderomotive effects are small. Even the absorption of a single XUV photon can trigger multi-electron dynamics, e.g. via Auger cascades~\cite{penent05}. Accordingly, XUV-pump XUV-probe experiments are a promising route to provide in-depth understanding of multi-electron dynamics, where one or more photons may be absorbed from both the pump and the probe lasers in order to produce multiply charged ions. Using state-of-the-art technology, such experiments can now be performed with attosecond to few-femtosecond time resolution. Indeed, two-photon, two-electron interactions have been observed with a temporal resolution down to 1.5~fs~\cite{tzallas11} and 500~as~\cite{takahashi13}.

The absorption of two XUV photons by an atom resulting in its double ionization may occur either in a direct or a sequential process. In the latter case, one electron is emitted following the absorption of a first XUV photon, and an intermediate ionic state is formed. Another electron is emitted when a second XUV photon is subsequently absorbed. Various experimental techniques have been applied to distinguish between direct and sequential XUV-induced processes in single-pulse experiments, including intensity-dependent studies~\cite{moshammer07} and the investigation of recoil-ion momentum distributions~\cite{rudenko08}. Previous experiments aimed at characterizing the XUV pulse duration used a pair of XUV pulses and relied on selecting suitable photon energies that favor direct two-photon-induced two-electron emission over the sequential absorption of two XUV photons~\cite{nabekawa05,mitzner09}.

\begin{figure*}[htb]
 \centering
  \includegraphics[width=17cm]{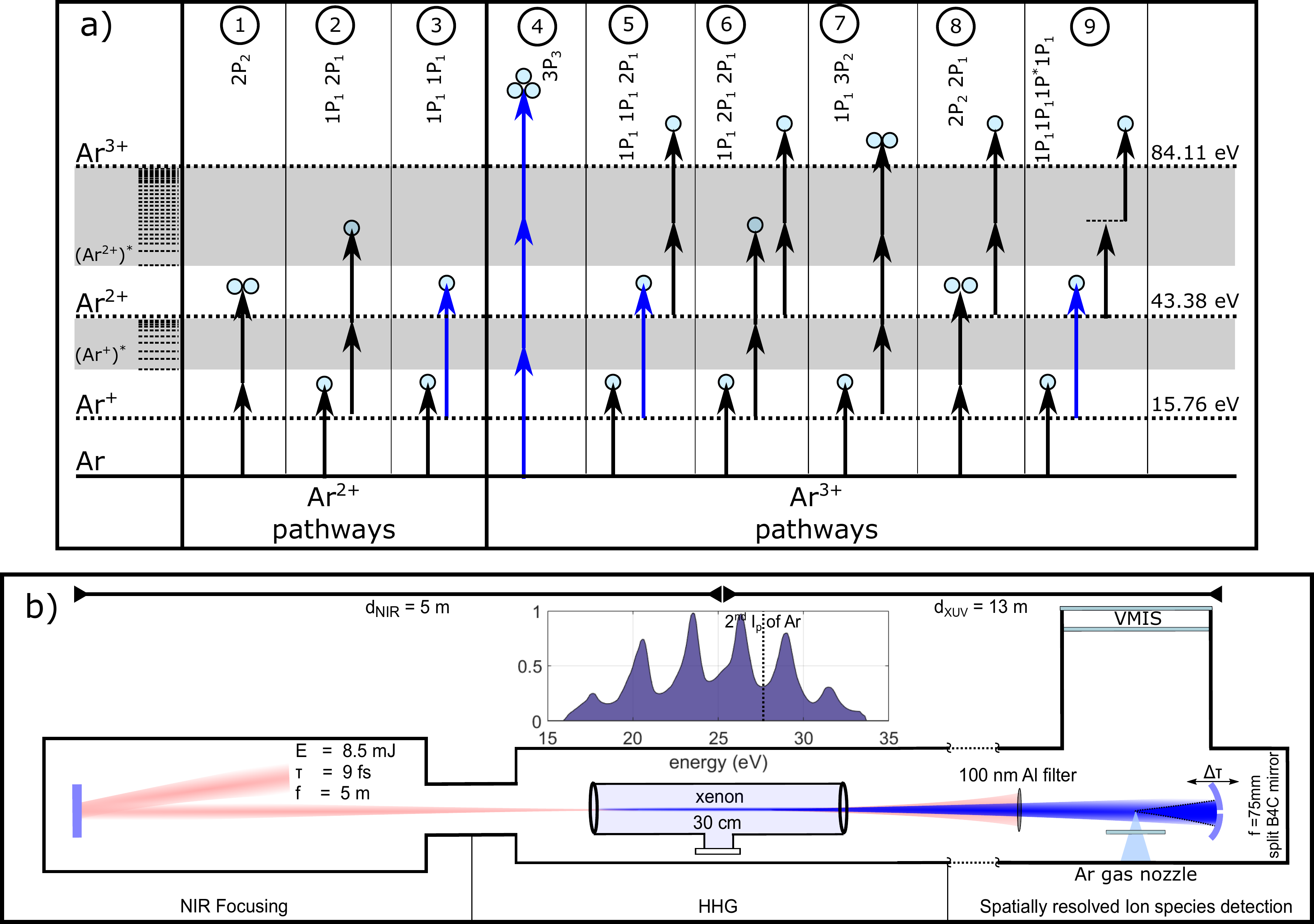}
 \caption{\label{figure_setup} \textbf{Multi-photon ionization scheme for Ar and experimental setup.} (a) Term scheme showing likely multi-photon ionization pathways leading to the generation of Ar$^{2+}$ and Ar$^{3+}$, including various direct and sequential ionization pathways (see main text for details). The dashed lines and the gray area indicate a high density of excited states. The notation $N\mathrm{P}_M$ is used to describe an $N$-photon absorption process leading to the emission of $M$ electrons. The black arrows correspond to XUV photons with an average photon energy of about 25~eV, whereas the blue arrows correspond to XUV photons that have a higher-than-average photon energy, which is required in some of the shown processes. (b) Experimental setup: NIR pulses are focused using a spherical mirror with a focal length of 5~m into a 30-cm-long gas cell filled with Xe. A split spherical B$_4$C-coated mirror is used to generate two replicas of the APTs and to focus them into the interaction region of a velocity-map imaging spectrometer (VMIS). The XUV photon spectrum obtained from a measured photoelectron spectrum is shown on top.}
\end{figure*}

The investigation of XUV-induced multi-photon multi-electron dynamics with attosecond time resolution has so far been hindered by the lack of suitable light sources. While extremely high XUV intensities can be obtained at free-electron lasers~\cite{sorokin07}, it has only recently become possible to generate attosecond pulses and attosecond pulse trains at these facilities~\cite{hartmann18, maroju20, duris20}. At the same time, attosecond pulses can be routinely generated using HHG~\cite{krausz09}. Recently, intense XUV pulses from HHG sources have been demonstrated, allowing the observation of multi-photon absorption. In Ar and Xe atoms, this has led to the observation of ion charge states up to Ar$^{5+}$~\cite{nayak18, senfftleben20, major21} and Xe$^{5+}$~\cite{bergues18}. In order to understand the underlying ionization mechanisms, attosecond-pump attosecond-probe spectroscopy represents a promising tool.

Here we present the results of an XUV-pump XUV-probe experiment in Ar using a pair of attosecond pulse trains (APTs) with an intensity envelope of $\approx 3$~fs~\cite{major20}, and covering a photon energy range from $16-34$~eV. Using photons in this photon energy range, Ar$^{2+}$ can be produced by both two- and three-photon processes, see Fig.~1(a). The absorption of two photons may take place in a direct (1) or sequential (3) two-photon process. In the first case, two electrons are emitted following the simultaneous absorption of two photons. In the second case, Ar$^{+}$ is formed after the absorption of a first XUV photon, followed by the absorption of a second XUV photon and the emission of a second electron. Ar$^{2+}$ may further be produced by three-photon absorption (2), involving a sequence of a one-photon step leading to Ar$^+$ and a two-photon process leading to Ar$^{2+}$. For the generation of Ar$^{3+}$, six likely pathways are shown, including a direct three-photon process (4) and a fully sequential pathway involving four single-photon absorption steps via intermediate states Ar$^+$, Ar$^{2+}$ and Ar$^{2+}$* (9). Note that a high density of excited states of Ar$^{2+}$ (indicated by the dashed lines and the gray area) lie within the photon energy range of the used APTs~\cite{NIST}. Pathways (5) - (8) involve sequences of one-, two-, and three-photon absorption. Further pathways are possible, but were not found to be important in this work.

To study the non-linear multi-photon ionization of argon with attosecond resolution, we applied a recently developed XUV intensity scaling scheme using an 18-m-long HHG beamline, see Fig.~1(b)~\cite{senfftleben20}. High harmonics were generated in a 30-cm-long gas cell filled with Xe using 9-fs-long NIR pulses centered around 800\,nm from an optical parametric chirped-pulse amplification (OPCPA) system~\cite{kretschmar20}. As shown in Fig.~1(b), two replicas of the APT were generated by a split-and-delay unit. The pulses were focused onto a jet of Ar atoms using a split B$_4$C-coated spherical mirror with a focal length of 75~mm. An XUV peak intensity of $1 \times 10^{14}$~W/cm$^2$ was estimated for each APT (see Methods). The generated ions were recorded using a velocity-map imaging spectrometer (VMIS)~\cite{eppink97} that was operated in spatial-map imaging mode~\cite{stei13}.

In what follows, we will present measurements of the Ar$^{2+}$ and Ar$^{3+}$ yields as a function of the delay between the two APTs. Our aim is to identify the dominant mechanisms responsible for the Ar$^{2+}$ and Ar$^{3+}$ formation (see Fig.~1(a)) on the basis of the measured delay dependence. In order to do so, we will compare the measured delay dependence with results of theoretical calculations including the mechanisms shown in Fig.~1(a). It should be emphasized that theoretical modeling of the experiment is extremely challenging, since the experiment collects signals from an extended focal volume where the two APTs overlap and interfere with each other, leading to a complicated XUV intensity and pulse shape distribution within the focal volume. Since, as a result, the computation of focal volume averaged results is unfeasible, we have first characterized the role of focal volume averaging in a simplified model, where we have calculated non-resonant two-photon ionization using XUV pulses obtained in a previous simulation using realistic NIR pulses~\cite{major20}. These calculations show that to a good approximation the experiment can be considered and modeled as an intensity autocorrelation (see Supplementary Information). Therefore, this approximation will be used within this paper.

\section*{Results}

\begin{figure}[htb]
 \centering
   \includegraphics[width=15cm]{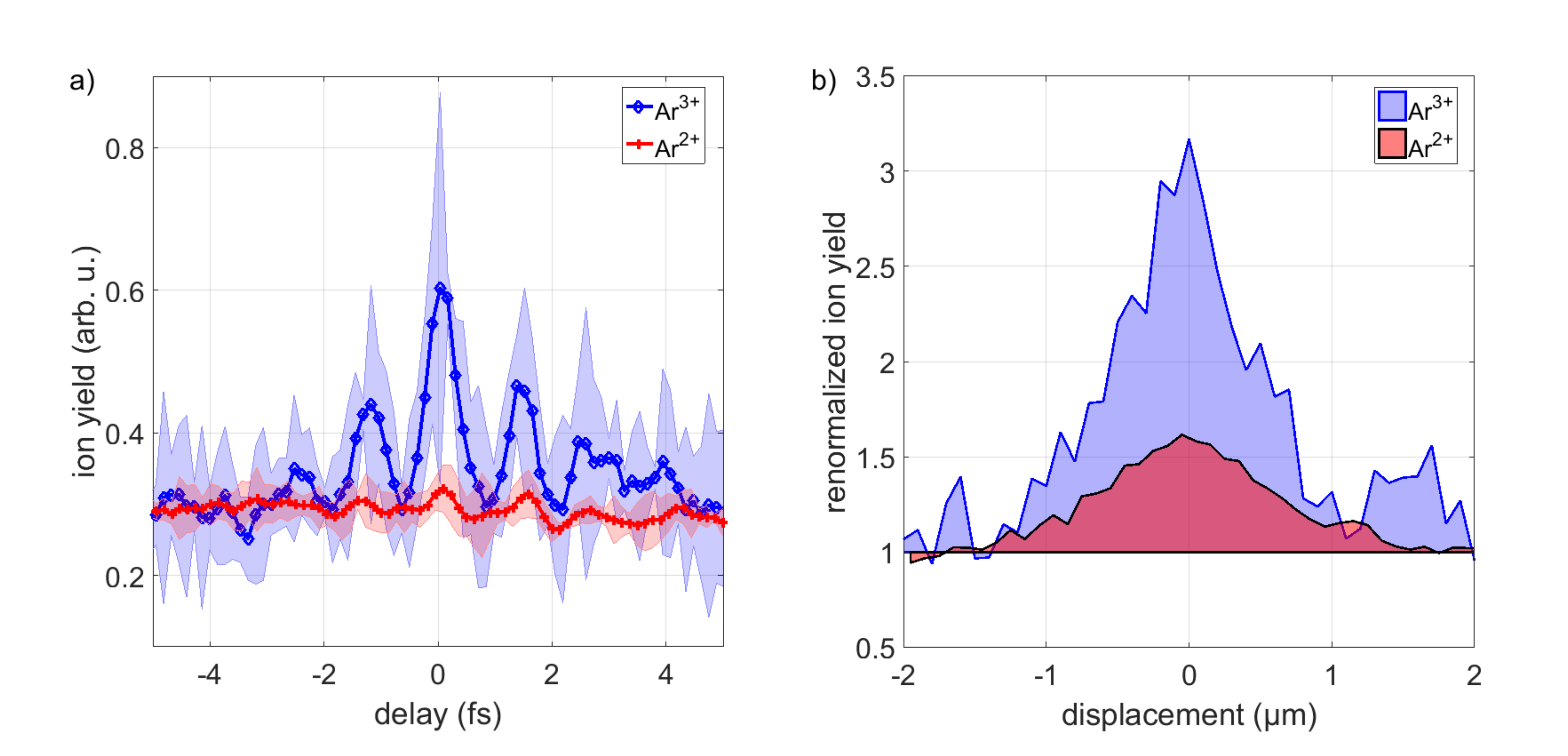}
 \caption{\label{figure_AC} \textbf{Measured Ar$^{2+}$ and Ar$^{3+}$ ion yields as a function of the XUV-XUV temporal and spatial overlap.} (a) The Ar$^{2+}$ ion yield (red curve) is only weakly modulated as a function of the XUV-XUV time delay, whereas clear oscillations with a period of 1.3~fs are observed in the delay-dependent Ar$^{3+}$ ion yield (blue curve). Error intervals given by the red- and blue-shaded areas are standard deviations obtained as the statistical error of four scans. Note that the absolute Ar$^{3+}$ yield is about two orders of magnitude smaller than the absolute Ar$^{2+}$ yield. (b) At a delay of 20~fs between the two APTs, the Ar$^{2+}$ ion yield (red curve) is enhanced by a factor of 1.6 when the XUV pulses spatially overlap, while an enhancement of 3.2 is observed for Ar$^{3+}$ (blue curve). The combination of these results show that the formation of Ar$^{2+}$ is dominated by a sequential pathway, whereas both sequential and direct processes are important for the generation of Ar$^{3+}$. }
\end{figure}

We have performed two types of measurements: In the first type of measurement, the two APTs were spatially overlapped, and the time delay between the two APTs was varied. The measured Ar$^{2+}$ and Ar$^{3+}$ ion yields as a function of the XUV-XUV time delay are presented in Fig.~2(a). The Ar$^{2+}$ yield (red curve) exhibits only weak oscillations with an oscillation amplitude of about 10~$\%$, whereas the Ar$^{3+}$ ion yield (blue curve) shows strong oscillations with an oscillation amplitude of about 2 as a function of the time delay between the two APTs. The period of the oscillations is about 1.3~fs, which corresponds to half the oscillation period of the NIR laser and suggests a strong enhancement of the Ar$^{3+}$ yield when individual attosecond pulses in the pump and probe APTs overlap. In the second type of measurement, the two APTs were delayed by 20~fs with respect to each other, and the focus position of one of the APTs was scanned in lateral direction with respect to that of the other. As shown in Fig.~2(b), the Ar$^{2+}$ ion yield was increased by a factor of 1.6 and the Ar$^{3+}$ ion yield was increased by a factor of 3.2 when the APTs spatially overlapped. Taken together, these experiments allow us to better understand the importance of both direct and sequential multi-photon processes.

In order to assess the relative importance of direct and sequential processes, we first of all make the observation (which is based on the afore-mentioned focal volume averaged simulations, see Sec.~II of Supplementary Information) that, except at zero delay, sequential two-photon two-electron ionization does not depend on the delay between the two APTs. In contrast, direct two-photon, two-electron ionization is about twice as efficient when the APTs temporally overlap as opposed to when they do not, since the peak intensity is twice as high when the two APTs overlap. We stress that this result is only obtained when realistic APTs resulting from simulations of our XUV source are used. It thus follows from the observed amplitudes of the oscillations in the Ar$^{2+}$ signal (Fig.~2(a)) that at least 10~$\%$ of the Ar$^{2+}$ yield is due to direct ionization, whereas up to 90~$\%$ is due to sequential processes. With regard to the second experiment, where the two APTs were delayed by 20~fs with respect to each other, we argue that in the absence of ground-state depletion (see Sec.~III of Supplementary Information) the yield of Ar$^{2+}$ resulting from direct ionization does not depend on whether or not the two APTs are spatially overlapped. By contrast, given that sequential ionization scales quadratically with the fluence, it is expected that the contribution from sequential ionization doubles when the two APTs spatially overlap. It follows that the 60~$\%$ increase in the Ar$^{2+}$ yield in the latter case suggests that at least 60~$\%$ of the Ar$^{2+}$ formation occurs by means of a sequential process and up to 40~$\%$ by a direct process. Combining both observations, we thus estimate that about 10-40~$\%$ of the Ar$^{2+}$ formation occurs by direct two-photon two-electron emission, and the remainder by sequential processes.   

The experimental results for Ar$^{3+}$ suggest that its formation takes place in a sequence of steps, of which at least one step involves the absorption of two or more XUV photons. Due to the large number and the complexity of the possible pathways, a definite assignment to one of the pathways shown in Fig.~1(a) is not possible. However, when one makes the assumption that Ar$^{3+}$ is generated via the Ar$^{2+}$ ground state (i.e. the first two XUV photons are preferentially absorbed sequentially), the most relevant pathways would be (5) and (9). The strong modulation of the Ar$^{3+}$ ion yield as a function of the time delay (Fig.~2(a)) would then mean that pathway (9) can be ruled out and that pathway (5) dominates. 

\begin{figure}[tb!]
 \centering
 \includegraphics[width=10cm]{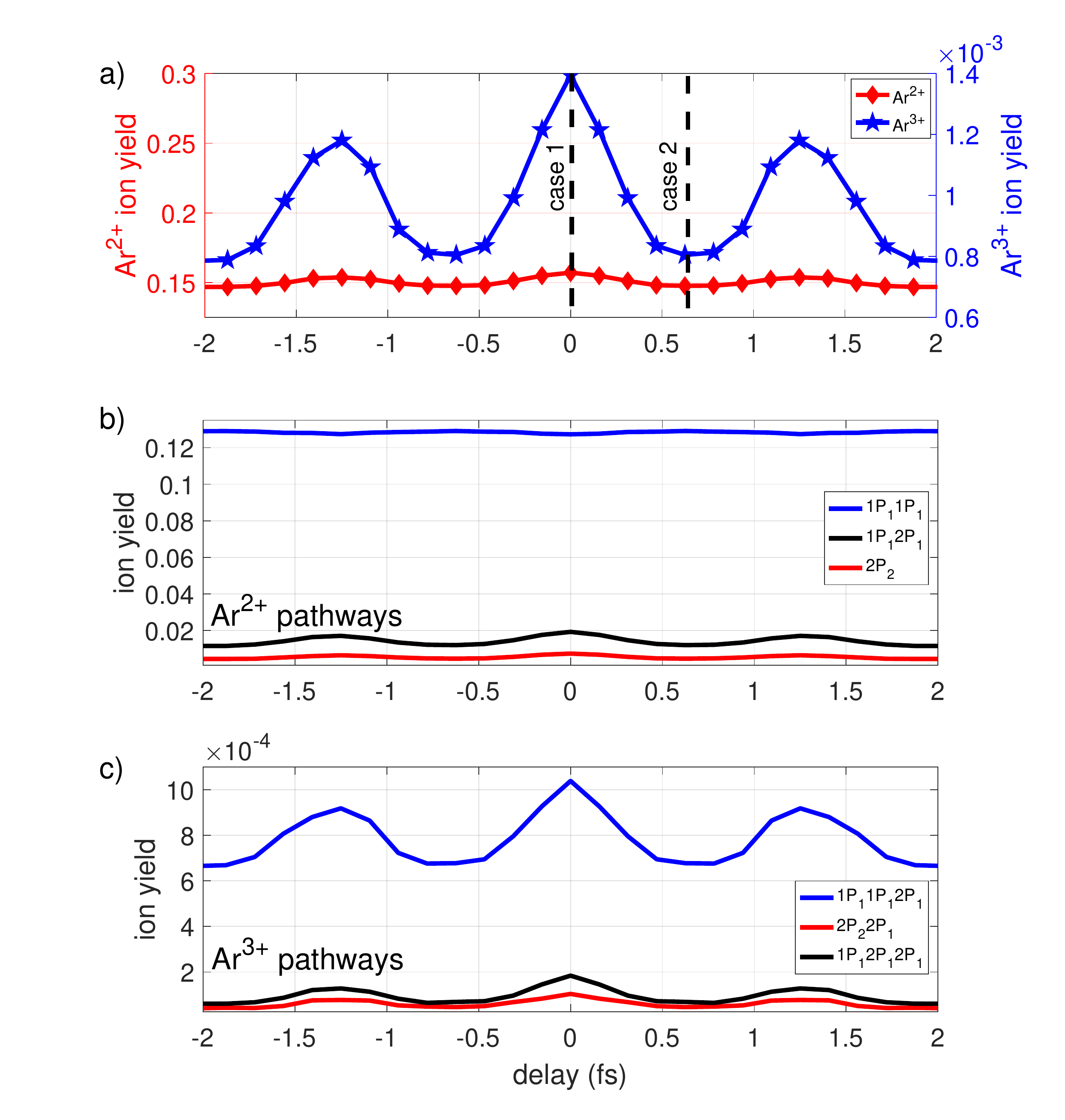}
 \caption{\label{figure_sim1} \textbf{Simulated Ar$^{2+}$ and Ar$^{3+}$ ion yields as a function of the XUV-XUV time delay.} (a) The simulated Ar$^{2+}$ ion yield (red curve) exhibits small delay-dependent changes of about 10~$\%$, while the relative Ar$^{3+}$ ion yield (blue curve) varies by a factor of almost 2. These results are in good agreement with the experimental results shown in Fig.~2. Case 1 and case 2 are two time delays for which the time-dependent ion formation is shown in Fig.~4. (b) Contributions to the Ar$^{2+}$ ion yield as a function of XUV-XUV time delay, dominated by sequential two-photon absorption (blue curve), which hardly shows any delay dependence. (c) The Ar$^{3+}$ ion yield (blue curve) is dominated by a sequence of two one-photon and one two-photon absorption steps (pathway (5) in Fig.~1(a)). All pathways contributing to the formation of Ar$^{3+}$ show clear oscillatory behavior as a function of the XUV-XUV time delay.}
\end{figure}

Deeper insights into the non-linear multi-photon ionization of Ar can be obtained by modeling, which will be described in the following. To be able to account for the broad bandwidth of the XUV pulses, a Monte Carlo technique was developed that computed the final ion yields and the pathways leading to the formation of the ions. Single-photon and direct two- and three-photon ionization processes were accounted for (see Methods and Supplementary Information for details). In the simulations, a pair of short APTs were used (see gray-shaded area in Fig.~4(a)), as previously obtained from HHG simulations~\cite{major20}. We note that it was not possible to implement the formation of excited ions in the model.

The simulated Ar$^{2+}$ and Ar$^{3+}$ ion yields as a function of the time delay between the two APTs are presented in Fig.~3(a). While the Ar$^{2+}$ ion yield exhibits weak oscillations with an amplitude of about 10~$\%$, the Ar$^{3+}$ ion yield strongly oscillates and is about twice as high at zero delay as compared to the case of non-overlapping pump and probe pulses. These results are in good agreement with the experimental observations (Fig.~2(a)). 

The contributions of different pathways leading to the generation of Ar$^{2+}$ and Ar$^{3+}$ are shown in Fig.~3(b)+(c). The formation of Ar$^{2+}$ (Fig.~3(b)) is dominated by a sequential pathway involving two single-photon absorption steps (pathway (3) (blue curve) in Fig.~1(a)), and shows almost no delay dependence. Further contributions stem from pathways (1) (red curve) and (2) (black curve), which involve a direct two-photon absorption step. These contributions are responsible for the weak oscillations observed in Fig.~3(a). The dominance of a two-photon sequential pathway as obtained from the simulations is consistent with the analysis of the experimental results. 

The generation of Ar$^{3+}$ (Fig.~3(c)) is found to be dominated by a three-step sequence of two one-photon and one two-photon absorption processes via intermediate states Ar$^+$ and Ar$^{2+}$ (blue curve, pathway (5) in Fig.~1(a)). Pathways (6) (black curve) and (8) (red curve) also contribute to the overall Ar$^{3+}$ ion yield. All these contributions exhibit clear oscillations as a function of the XUV-XUV time delay. We note that in our simplified model, we found that a sequence of three ionization steps results in an increased ion yield by a factor of 4, when the two APTs are spatially, but not temporally overlapped (see Supplementary Information). Therefore, we note that the experimental observations, which showed an increase of the Ar$^{3+}$ ion yield by a factor of 3.2 at spatial overlap and an additional increase of a factor 2 at temporal overlap between the two APTs, are consistent with pathway (5) playing an important role for the formation of Ar$^{3+}$, as predicted by the Monte Carlo simulations. 

\begin{figure}[tb]
 \centering
  \includegraphics[width=11cm]{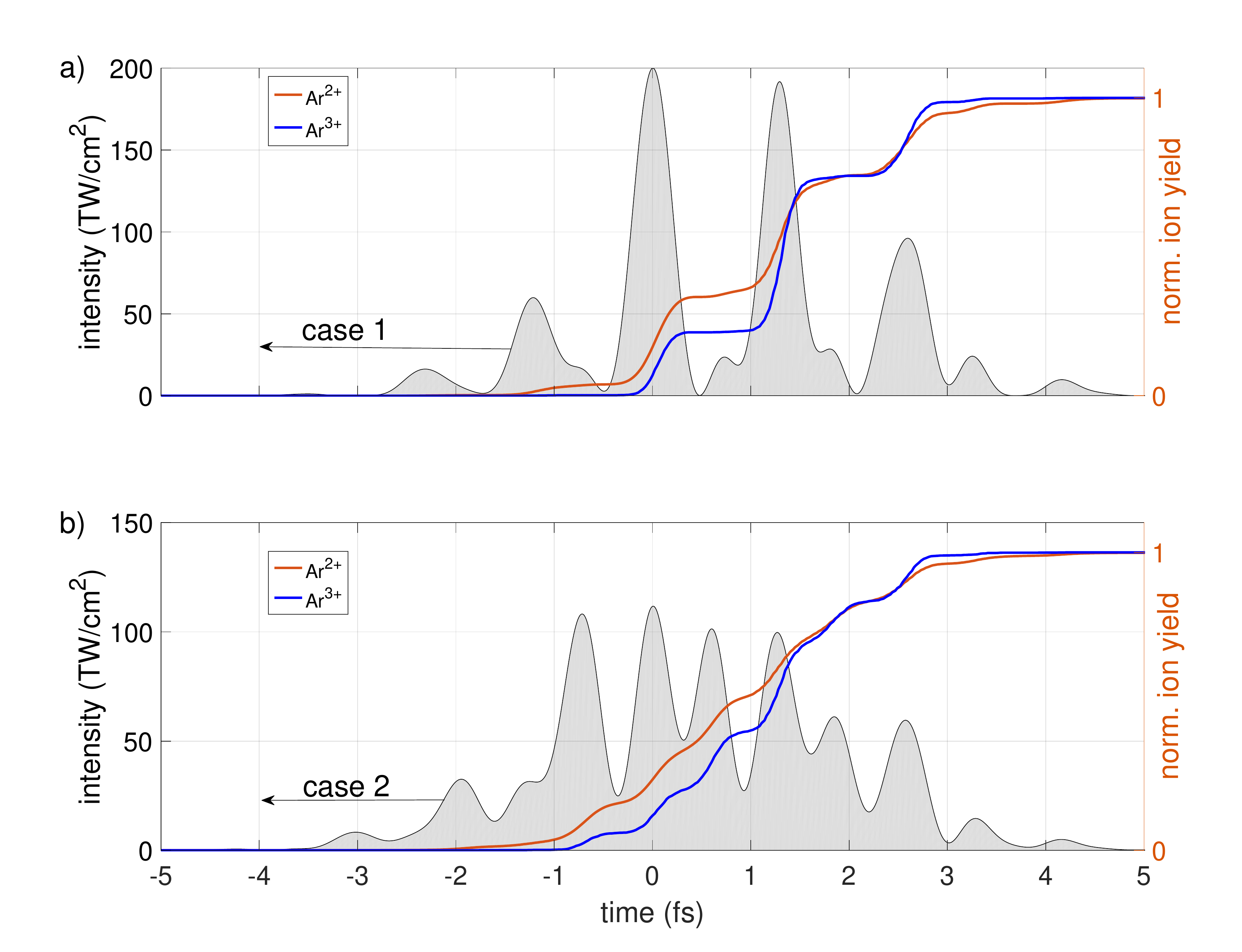}
 \caption{\label{figure_time-dependence} \textbf{Simulated evolution of the Ar$^{2+}$ and Ar$^{3+}$ ion yields in time.} The results are normalized with respect to the final ion yields and are shown for XUV-XUV delays of (a) 0~fs (case 1 in Fig.~3) and (b) 0.66~fs (case 2 in Fig.~3), see the gray-shaded areas for the corresponding effective XUV pulses. In both cases, ionization occurs in a stepwise fashion due to the APT structure.}
\end{figure}

The numerical simulations provide information on the formation of Ar$^{2+}$ and Ar$^{3+}$ ions in the time domain. The corresponding results are presented in Fig.~4 for XUV-XUV time delays of (a) 0~fs and (b) 0.66~fs (see the gray-shaded areas for the corresponding effective XUV pulse structures). As a result of the APT structure, the ion formation occurs in steps in both cases. The increase of the Ar$^{3+}$ ion yield in Fig.~4(a) is particularly large in the temporal window between 1.1 and 1.6~fs. This is a consequence of the fact that the Ar$^{3+}$ ion formation depends both on the instantaneous Ar$^{2+}$ ion population and the instantaneous XUV intensity.

\section*{Summary and Outlook}

Our results demonstrate that the application of intense attosecond pulse trains opens the way to better understand complex multi-photon multi-electron dynamics. In our study, the combination of temporally and spatially resolved measurements applied to different ionic species made it possible to separately study the role of direct and sequential multi-photon processes. Therefore, the demonstrated method has the potential to investigate in detail the complex and highly non-linear multi-photon ionization of atoms and molecules observed at free-electron lasers, which resulted in charge states up to Xe$^{48+}$~\cite{sorokin07, rudek12, rudenko17}, Ne$^{10+}$~\cite{young10} and I$^{47+}$~\cite{rudenko17}.

The high XUV intensities obtained here furthermore pave the way for attosecond-pump attosecond-probe spectroscopy at higher XUV photon energies, a regime that is typically difficult to access because of low HHG conversion efficiencies. This could make it possible to study multi-electron dynamics and electron-electron correlation following the removal of core electrons in atoms and molecules with attosecond resolution. Examples include the study of Auger cascades and double Auger decay processes, in which the relaxation of a valence-shell electron to an inner-shell vacancy leads to the sequential or simultaneous emission of two Auger electrons~\cite{penent05}.

\section*{Acknowledgments}
We thank Valer Tosa for providing his HHG simulation code. A. E. would like to thank Peter Lambropoulos for useful discussions. A. E. further acknowledges the use of the Legion computational resources at UCL. This work was funded by the Leverhulme Trust Research Project Grant No. 2017-376. We acknowledge KIF\"U for awarding us access to resources based in Hungary in Debrecen and Szeged. The ELI-ALPS project (GINOP-2.3.6-15-2015-00001) is supported by the European Union and co-financed by the European Regional Development Fund.

\vspace{1cm}
\noindent
\textbf{Experimental methods}\\
In the experiments, 8.5~mJ NIR pulses were used, and the pulse energy was further reduced to 6~mJ during the HHG optimization process using a motorized iris. Rather than maximizing the XUV pulse energy as is done in most loose-focusing HHG setups, an XUV intensity scaling scheme was applied which optimizes the XUV peak intensity by reducing the XUV waist radius~\cite{senfftleben20}. Considering that the total beamline length was 18~m, a spherical mirror with a relatively short NIR focal length of 5~m was used, and the distance from the NIR focal plane to the XUV focusing mirror was 13~m, leading to a large demagnification of the XUV beam size~\cite{senfftleben20}. High harmonics were generated in a 30-cm-long gas cell filled with Xe, where NIR propagation effects were previously shown to play an important role~\cite{major20}. A 100-nm-thick Al filter was used to block the fundamental light. The XUV beam was split and delayed using two sections of a spherical mirror with a focal length of 75~mm and a boron carbide (B$_4$C) coating. In the current experiments, we estimate that each XUV focusing mirror reflected about 40~$\%$ of the XUV beam, leading to a beam waist radius that was estimated to be 1.3~$\mu$m in the vertical direction~\cite{senfftleben20} and $2.5 \times 1.3$~$\mu$m$=3.25$~$\mu$m in the horizontal direction. The width of the Ar$^{2+}$ ion yield distribution as a function of the spatial displacement of one of the APTs (Fig.~2(b)) is consistent with the estimation of the XUV beam waist radius. Assuming an XUV pulse energy of 10~nJ that was reflected from each mirror and using the simulated APT structure (see Fig.~4)~\cite{major20}, the XUV peak intensity of each pulse was estimated as $1\times 10^{14}$~W/cm$^2$. Ar atoms were injected by a piezoelectric valve~\cite{irimia09}, and the central part of the atomic beam was selected by a molecular beam skimmer with an orifice diameter of 0.5~mm. Ions in different charge states were recorded using a velocity-map imaging spectrometer~\cite{eppink97} that was operated in spatial-map imaging mode~\cite{stei13}. The ion yields of individual ion charge states were recorded by gating the microchannel plate / phosphor screen detector.

\vspace{1cm}
\noindent
\textbf{Numerical methods}\\
A Monte-Carlo technique~\cite{robert13} was employed to compute the ion yields and the pathways leading to the formation of the ions. Namely, starting from an initial state, we accessed only one energetically allowed state at a time, which is determined in a stochastic manner~\cite{rudek12, son12, jurek16}. Moreover, due to the finite bandwidth of the XUV pulses, at each time step of the propagation, using a method referred to as importance sampling~\cite{rubinstein16}, one, two, and three photons in the energy interval from 15.9~eV to 33.7~eV were selected to account for the respective multi-photon process. The importance sampling distribution is given by the experimentally determined spectrum of the XUV pulse. To compute single-photon cross sections, a formalism previously developed to describe single-photon ionization processes in argon was used~\cite{wallis15}. Employing a scaling method described in Ref.~\cite{lambropoulos87}, two- and three-photon cross sections for argon were estimated as 10$^{-50}$~cm$^4$s and 10$^{-84}$~cm$^6$s$^2$, respectively. These approximate values are in agreement with two- and three-photon cross sections previously obtained to describe the interaction of XUV pulses with argon~\cite{karamatskos13} and neon~\cite{nikolopoulos14}. A more detailed description of the numerical methods is given in the Supplementary Information.

\bibliography{Bibliography}
\end{document}